**Title:** The role of plasma-activated water on the growth of freshwater algae *Chlorella Pyrenoidosa* and *Chlorella Sorokiniana*

**Authors**

Vikas Rathore[1,2*] and Sudhir Kumar Nema[1,2]

1. Atmospheric Plasma Division, Institute for Plasma Research (IPR), Gandhinagar, Gujarat 382428, India

2. Homi Bhabha National Institute, Training School Complex, Anushaktinagar, Mumbai 400094, India

*Email: vikas.rathore@ipr.res.in

**Abstract**

In the present work we have conducted two studies. In the first study, we investigated the role of plasma-activated water (PAW) in algae growth inhibition and in the second study, efforts are made to understand the role of PAW as a nitrogen source for algae growth enhancement. Two freshwater algae species are selected for the present study named *Chlorella Pyrenoidosa* and *Chlorella Sorokiniana*. The PAW is prepared using a pencil plasma jet and air as a plasma forming gas. The plasma is characterized electrically and identification of generated species in plasma is carried out using optical emission spectroscopy.

    The study clearly indicated that more oxidizing PAW exhibits algicidal effect. The PAW treatment with both the algae species substantially decreased their growth compared to control. Moreover, the morphology of algae cells showed damage and cells structure get ruptured after PAW treatment.

In the second study, a less reactive PAW (low oxidizing potential) used as a nitrogen replacement in Bold's Basal Medium. The PAW-grown *Chlorella Pyrenoidosa* and *Chlorella Sorokiniana* showed higher growth compared to control. Also, a higher concentration of chlorophyll 'a' and 'b', sugar, and protein observed compared to control. Further, we observed lower antioxidant enzymatic activities in PAW-grown algae compared to control. In conclusion, the PAW has algicidal efficacy as well as can be used as nitrogen source in aquaculture to enhance algae growth.

**Keywords:** plasma activated water, reactive oxygen-nitrogen species, plasma characterization, algicidal efficacy, algae growth enhancement

**1. Introduction**

Algae are well known for their enormous potential to be used in numerous applications due to their higher growth rate and economic viability. These applications include biofuel synthesis and bioenergy production, as a food for animals (including marine animals) and humans, as medicine, used as vitamins and antioxidant supplements, as natural pigments and dyes, and for wastewater treatment, etc.(1-4). In addition, algae play a significant role in environmental sustainability. It utilizes carbon dioxide ($CO_2$) for its growth through the process of photosynthesis, hence contributing to $CO_2$ biofixation. In addition, it is a well-known bio-sorbent meaning it has the affinity to remove various contaminants from wastewater such as heavy metals, antibiotics, and industrial waste including pharma industries, food industries, and dairy industries, etc(1). Along with this, it has been used in industries like agriculture, domestic, and pharmaceutical to reduce toxicity levels in industrial effluent. It can uptake various nutrients present in those waste and utilize it for its growth and development. In conclusion, it captures $CO_2$ from the environment and removes contaminants from wastewater for its growth, and generates oxygen through photosynthesis. The algae can be easily harvested from various water bodies without additional investment. Hence, algae cultivation is a low investment high throughput industry. The produced algae biomass can be used to extract lipid which can be converted to biodiesel, a renewable source of energy. Moreover, it has the affinity to convert energy composed in sunlight to various products like protein, carbohydrate, and lipid, etc. through the process of photosynthesis(1).

As discussed above, algae have a high growth rate and can be grown in places that have water bodies which create problems like an algae bloom, the rapid growth of algae in cultivation tanks, and plant growing factories (5-7). It normally occurs due to the presence of excess nutrients such as nitrogen and phosphorus in water. Some of the problems associated with the algae bloom are reducing $O_2$ levels in water and preventing sunlight from reaching the water organism since it covers the water surface. In addition, some algae contain toxins which can result in various health issues in animals while consuming algae or drinking water contains toxins and that leads to threats to aquatic life. Such algae bloom is known as harmful algae bloom (6). The cleaning of algae from cultivation tanks and various apparatus in plant growing factories created an additional economic burden in the form of labor and cleaning reagents costs.

The present work discusses the role of plasma activated water (PAW) in algae growth enhancement as well as inhibition. The plasma exposure to water chemically modified the water due to the formation of various reactive oxygen-nitrogen species in it. The plasma is an ionized state of matter which contains radiation, high-energy electrons, ions, atoms, and neutrals. These excited states of plasma radicals and species when exposed to water formed stable RONS in it. The existence of RONS makes the water active and this water known as plasma activated water (PAW). This PAW contains species like $NO_3^-$ ions, $NO_2^-$ ions, $O_3$ (aq.), $H_2O_2$, $ONOO^-$, $\cdot OH$ (aq.), and NO (aq.), etc.(8-13) The existence of these RONS in PAW makes it a fruitful product that could be used in numerous applications. The presence of reactive oxygen species ($O_3$ (aq.), $H_2O_2$, $ONOO^-$, and $\cdot OH$ (aq.), etc.) in PAW supports applications like microbial inactivation (bacteria, fungi, virus, and pest), food preservation (fruits, vegetables, meat products, and seafood, etc.), disinfection of medical devices, and selective killing of cancer cell, etc.(14-21) Moreover, the reactive nitrogen species ($NO_3^-$ ions and $NO_2^-$ ions, etc.) play a significant role in the agriculture field as a nitrogen fertilizer to enhance seeds germination and plant growth, nitrogenous acids, etc(22-25).

The work regarding the role of PAW in algae growth is very limiting. Previously, Sukhani et al.(2, 4) used PAW as a nitrogen source in algae growth enhancement. They used a mixotrophic culture of algae (collected from sewage water) for the study. Also, the reported work showed only primitive investigation such as the effect of algae biomass yield and chlorophyll concentration. Moreover, commenting on mixed algae culture response and biochemistry towards PAW is very difficult, Since, mixed culture contains a lot of unidentified algae species along with unknown foreign species. In addition, other growth attributes such as variation in chlorophyll 'a', and 'b', carotenoids, soluble sugar and protein, oxidative stress, and enzymatic activities, etc. have not been reported. The algicidal efficacy of plasma-activated water was previously reported by Mizoi et al. (5). They showed pH of PAW contributes to algae (*Chlorella vulgaris*) growth inhibition. They imply the ONOOH present in PAW at low pH 2.2 penetration the algae membrane and destroy it that lead to algae growth inhibition. Moreover, morphology changes in cells after PAW have not been reported. These are the only work reported as per the best of the authors' knowledge that signifies the role of PAW in algae growth enhancement as well as in growth inhibition. Hence, the algicidal efficacy of PAW, as well as its uses as a nitrogen source in algae growth enhancement, has to be explored in a systematic and detailed manner. It becomes the motivation of the present work.

The current manuscript tries to overcome this research disparity. Hence for the same two freshwater unicellular algae *Chlorella* species named *Chlorella Pyrenoidosa* and *Chlorella*

*Sorokiniana* are chosen. The *Chlorella Pyrenoidosa* and *Chlorella Sorokiniana* have high lipid productivity (26, 27). Hence, plays a significant role in green energy in the form of biodiesel production, a step towards sustainable development. Moreover, they can utilize organic and inorganic nutrients in wastewater for their growth (26, 27). As a result, they also reduce the inorganic and organic pollutant levels in wastewater. Along with that, *Chlorella Pyrenoidosa* and *Chlorella Sorokiniana* have applications in the field of biofertilizers, improving biofuel efficiency, medicine, and food supplements, etc.(28-31)

The present work discussed the role of PAW in algae growth inhibition by analyzing the algae growth after PAW treatment. Moreover, the damaged algae cells' morphology is analyzed using scanning electron microscopy. The second part of the present work discussed the role of PAW as a nitrogen replacement in algae growth enhancement. Along with algae growth study, the grown algal is characterized by studying chlorophyll 'a' and 'b' and carotenoids in cells. The change in the nutritional value of algae is studied by analyzing soluble sugar and soluble protein. The oxidative stress created in algae cells during growth is studied by analysis of $H_2O_2$ concentration, and electrolytic and phenolic leakages. At last, the various antioxidant enzyme activity is studied. Since, the enhanced level of oxidative stress in cells enhances the antioxidant enzyme activity to prevent the cells' damage (32-34).

## 2. Material and Methods

### 2.1 Electrical and emission optical characterization of DBD-PPJ

A dielectric barrier discharge (DBD) pencil plasma jet (PPJ) was used for plasma activated water (PAW) preparation. Figure 1 (a) showed schematic of PAW production along with plasma diagnosis (electrical and optical emission characterization). The DBD-PPJ setup consists of a coaxial cylindrical assembly of ground and power electrodes separated by a quartz tube as a dielectric. More detail about the DBD-PPJ setup can be found in our previously reported work(9, 15). This DBD-PPJ setup was powered with high voltage high frequency alternating current (AC) power supply (power rating 10 kVA, voltage: 0-10 kV, current: 0-999 mA, frequency: 0-40 kHz) (figure 1 (a)). The power supply was operated in current limiter mode which prevented the current shoot up during DBD discharge. All experiments were performed at a constant frequency of 40 kHz.

The voltage drop in the DBD-PPJ setup was evaluated using a 1000x probe (Tektronix P6015A) and a 4-channel, 100 MHz bandwidth, 2 GS s$^{-1}$ sampling rate digital oscilloscope (Tektronix TDS2014C) as shown in figure 1 (a). The total current (conduction current and discharge current) and transported charge were evaluated by measuring a voltage drop across 30-ohm non-inductive resistor and 100 nF non-polarized capacitor in series with the ground using a 10x (Tektronix TPP0201) voltage probe and oscilloscope (figure 1 (a)).

The air emission spectrum in plasma afterglow region was measured by capturing light photons from DBD-PPJ air discharge using an optical fiber and a spectrometer (UVH-1 – compact spectrometer) within the spectral range of 290 nm and 925 nm (figure 1 (a)).

## 2.2 PAW production and measurement of physicochemical properties and RONS concentration

A fixed volume (20 ml for growth inhibition and 200 ml for growth enhancement) of water (ultrapure milli-Q water or demineralized water (DM water)) was kept in 600 ml of a glass beaker. The distance between the water surface DBD-PPJ lower edge was kept fixed at 10 mm. The water was exposed to plasma for 30 min of time duration. Air was fed from top to DBD-PPJ setup using an air pump. The airflow rate was kept constant at 10 l min$^{-1}$ using an air rotameter. To enhance the solubilities of generated plasma species in the water, stirring of water (using magnetic stirrer) was performed during plasma-water interaction. In addition, 600 ml of the beaker was positioned in a bowl having an ice-water mixture so the PAW temperature was maintained at 0 °C (measured using laser-guided infrared temperature gun (Helect)) during plasma-water exposure (schematic and picture of the same shown in supplementary figure S1).

The plasma-water exposure physicochemically modified the water properties. The physicochemical changes that occur in water are electrical conductivity (EC), pH, total dissolved solids (TDS), and oxidation-reduction potential (ORP) were measured. Following instruments were measured to measure these properties: pH meter (Hanna Instruments HI98121), ORP meter (HM Digital ORP-200), and TDS meter (HM Digital AP1), and EC meter (HM Digital COM-100).

The dissolved reactive species in water due to plasma-water exposure were measured semi-quantitatively and quantitatively. The measured reactive oxygen-nitrogen species (RONS) in PAW were $NO_2^-$ ions, $NO_3^-$ ions, $H_2O_2$, and dissolved $O_3$, etc. The RONS semi-quantitative evaluation in PAW was performed using test strips and colorimetry test kits. The $NO_2^-$ ions and $H_2O_2$ concentrations were semi-quantitatively determine using test strips (QUANTOFIX nitrite and QUANTOFIX peroxide 25). The $NO_3^-$ ions and dissolved $O_3$ concentrations were semi-quantitatively determine using colorimetry test kits (VISOCOLOR Nitrate and HI-38054 Ozone test kit).

The quantitative determination of RONS concentration in PAW was performed using UV-visible spectroscopy (SHIMADZU UV-2600). A standard curve of RONS such as $NO_2^-$ ions ($\lambda_{max}$ = 540 nm), $NO_3^-$ ions ($\lambda_{max}$ = 220 nm), and $H_2O_2$ ($\lambda_{max}$ = 407 nm) prepared using $NaNO_2$ salt, $NaNO_3$ salt, and 30% $H_2O_2$ solution. The molar attenuation coefficient of $NO_2^-$ ions, $NO_3^-$ ions, and $H_2O_2$ obtained for the standard curve using Beer's Lambert law were given as 0.0009 $(\mu g\ l^{-1})^{-1}\ cm^{-1}$, 0.0602 $(mg\ l^{-1})^{-1}\ cm^{-1}$, and 0.4857 $(mM)^{-1}\ cm^{-1}$ respectively. The dissolved $O_3$ present in PAW was determined using the indigo colorimetry volumetric method ($\lambda_{max}$ = 600 nm)(9, 15). The expression is given as:

$$\frac{mg}{l} \text{ of } O_3 = \frac{100 \times \Delta A}{f \times b \times V} \tag{1}$$

Where *ΔA* is difference between absorbance at 600 nm between control (black) and sample, *f* is sensitivity factor (0.42), and *V* is volume of sample in milliliters.

All the chemicals and reagent used in present study are of high purity and laboratory standard purchase from Sigma-Aldrich (MERCK), Hi-media, Sisco Research Laboratories (SRL chemicals) Pvt Ltd, and High Purity Laboratory Chemicals (HPLC) Pvt. Ltd., etc.

## 2.3 Preparation of algal culture

The pure algal species of *Chlorella Pyrenoidosa* (NCIM Accession No. 2738) and *Chlorella Sorokiniana* (NCIM Accession No. 5673) were obtained from National Chemical Laboratory, Pune, India. The procured algae species were cultured in manually prepared sterile Bold's Basal Medium (BB Medium) till the growth inhibition and enhancement study was performed.

## 2.4 Algae growth inhibition study

Figure 2 showed the schematic of the algae growth inhibition study using PAW and control (ultrapure milli-Q water or demineralized (DM) water). A 0.1 ml of algae was added to 0.9 ml of PAW or control. The PAW or control treatment time with algae varied from 30 min to 240 min. Once the PAW or control treatment time was over, the 1 ml of sample (PAW or control + algae) was transferred to 200 ml of BB Medium that was kept in 250 ml of Erlenmeyer flask. The flask was placed over a rotary shaker at a constant rpm of value 80 rpm. The air was fed to the flask to provide $CO_2$ to algae using an air pump. The growth of algae was monitored regularly up to 15 days by measuring absorbance at 680 nm (temperature ~ 25 °C, relative humidity ~ 40, and an on-off light cycle (irradiance 44 W m$^{-2}$) 16-8 hours). The picture of the PAW and control grown algae is also shown in figure 2.

### 2.4.1 Morphology analysis of PAW treated algae

For morphology analysis, the PAW and control treated cells were centrifuged and the supernatant was discarded to obtain a dense pellet. The cells were prefixed using a 6% glutaraldehyde solution for 60 min. The cells were washed with phosphate buffer saline (PBS) solution three times to remove excess glutaraldehyde. Dehydration of cells were performed using different concentration of ethanol (35%, 50%, 75%, 95%, and 100% (absolute)) with 10 min of treatment time. The excess ethanol was removed by washing with PBS solution at least three times. A 1% osmium tetroxide solution was used to post-fixing of cells for 60 min duration. Then, the cells were washed with ultrapure milli-Q water and excess osmium tetroxide was removed by centrifuge (10,000 rpm for 10 min) and discarded the supernatant. At last, the cells were mixed with hexamethyldisilazane (HMDS) to achieve cell drying to a critical point. The HMDS mixed cells were transferred to conducting silicon wafer and left overnight in a desiccator to get air dry(35).

The silicon wafer carrying cells were mounted on the mono high-resolution scanning electron microscope (HR-SEM) sample holder with the help of double-sided adhesive carbon tape. To make the cells conducting, the cells were sputtered with gold nano-particles (Quorum Q150R ES) for 60 seconds. The cells HR-SEM (Zeiss MERLIN Series) images were taken in secondary electron mode with 20.00 KX (20,000) magnification with electron high tension voltage of 5.0 kV.

**2.5 Algae growth enhancement study**

The schematic of algae growth enhancement is shown in figure 2. In which PAW was used as a nitrogen replacement in BB medium (for growth enhancement study PAW terminology signifies BB medium – nitrogen salts + PAW as nitrogen source). The algae growth using PAW as medium (BB medium – nitrogen salts + PAW as nitrogen source) was compared with BB medium (positive control) and BB medium without nitrogen (negative control). A 0.1 ml of algae culture each was added to 200 ml of PAW, negative control, and positive control were kept in 250 ml of Erlenmeyer flask. The algae culture containing flask was placed over a rotary shaker rotating at a fixed rpm of 80 rpm (temperature ~ 25 °C, relative humidity ~ 40, and an on-off light cycle (irradiance 44 W m$^{-2}$) 16-8 hours). The enhancement of algae growth was monitored regularly up to 15 days by measuring optical density at 680 nm using a UV-visible spectrophotometer.

### 2.5.1 Measurement of algae chlorophyll and carotenoids

A 10 ml of grown algae culture was centrifuged at 6500 rpm for 15 min to obtain dense algae pellets and the supernatant was discarded. A 1 ml of chilled 80% acetone was added to a freshly prepared algae pellet. The algae acetone mixture was vortexed vigorously so enclosed chlorophyll and carotenoids in the cell membrane came out and mixed with acetone. Then the sample was centrifuged at 12,000 rpm for 10 min and collected the supernatant. The chlorophyll and carotenoids were measured in collected supernatants using a UV-visible spectrophotometer. The expression used to calculate chlorophyll and carotenoids (36) is given as:

Chlorophyll 'a' $\quad\quad C_a\left(\frac{\mu g}{ml}\right) = 12.25 A_{663} - 2.79 A_{647}$ (2)

Chlorophyll 'b' $\quad\quad C_b\left(\frac{\mu g}{ml}\right) = 21.5 A_{647} - 5.1 A_{663}$ (3)

Carotenoids $\quad\quad C_{x+c}\left(\frac{\mu g}{ml}\right) = (1000 A_{470} - 1.82 C_a - 85.02 C_b)/198$ (4)

### 2.5.2 Measurement of sugar and protein in algae

To determine the soluble protein and total sugar concentration, a 20 ml algae culture was centrifuged and the supernatant was discarded and collected as a dense pellet. For soluble sugar, 1 ml of 80 °C temperature ethanol 80% volume/volume (warm solution) was added to the algae pellet and the solution was vortexed. Then the solution was incubated for 10 min in a boiling water bath. So, the bound soluble sugar in algae cells comes out and was dissolved in ethanol. The solution was cooled to room temperature and centrifuge at 12,000 rpm for 10 min and collected the supernatant for soluble sugar analysis. The collected supernatant was mixed with 0.2% freshly prepared ice-cooled anthrone reagent. The developed reaction product color due to the addition of anthrone reagent showed maximum absorbance at 590 nm (37). To measure the unknown soluble sugar concentration extracted from algae cells a standard curve was prepared. The standard curve (shown in figure S2 of supplementary material) was made using different concentrations of dextrose as soluble sugar. The molar extinction coefficient obtained from standard sugar solution was given as 0.0064 $(mg\ l^{-1})^{-1}\ cm^{-1}$.

For soluble protein, 1 ml of ice-cooled extraction buffer was added to the collected algae pellet. The extraction buffer (lysis buffer) was prepared using 0.07% β-mercaptoethanol (β-ME) and 10% trichloroacetic (TCA) acid in 1 ml of acetone solution. The extract buffer lysed the cells and the soluble protein dissolved in the extraction buffer. The extraction buffer carrying cells was centrifuged at 12,000 rpm for 10 min and the supernatant was collected for soluble protein analysis. All protein extraction experiments were performed at ~ 4 °C to prevent denaturation of protein. In 0.1 ml of collected supernatant, 0.1 ml of 2 N sodium hydroxide was added and hydrolyzed the solution in a boiling water bath for 10 min. Then the hydrolyzed solution was cooled down to room temperature and 1 ml of freshly prepared complex reagent was added. The complex reagent was prepared using 2% sodium carbonate ($Na_2CO_3$), 1% copper sulfate pentahydrate ($CuSO_4.5H_2O$), and 2% potassium sodium tartrate ($KNaC_4H_4O_6.4H_2O$) in the composition as follows 100:1:1. After adding the complex reagent, the solution was incubated for 10 min at room temperature. At last, a 0.1 ml of folin reagent was added and left the solution in dark for 30 to 60 min for reaction product color development(38). The developed color showed maximum absorbance at 750 nm. To determine unknown soluble protein concentration, a standard curve of protein solution of different concentrations was prepared using bovine serum albumin as standard protein. The molar extinction coefficient obtained from the standard curve was 0.1 $(\mu g\, l^{-1})^{-1}\, cm^{-1}$.

**2.5.3 Determination of $H_2O_2$ concentration, electrolytic and phenolic leakage from algae cells**

A fresh prepared algae pellet as discussed above used for $H_2O_2$ determination in algae cells. For $H_2O_2$ extraction, 1 ml of 0.1% TCA was added to cells and vortexed cells containing samples. The sample was centrifuged at 12,000 rpm for 10 min and the supernatant was collected for $H_2O_2$ analysis. In 0.25 ml of supernatant, a 0.25 ml of 100mM phosphate buffer, and 1 ml of 1M potassium iodide reagent were added. The final solution was incubated in dark for 60 min for reaction product color development. This color complex showed maximum absorption at 390 nm(39). The unknown concentration of $H_2O_2$ in algae was quantified using a standard curve. The standard curve of $H_2O_2$ was prepared using a different concentration of 30% $H_2O_2$ solution. The obtained molar extinction coefficient from the $H_2O_2$ standard curve was given as 0.9 $\mu M^{-1}\, cm^{-1}$.

For electrolytic and phenolic leakage, a freshly prepared algae pellet was dissolved in 20 ml of ultrapure milli-Q water. The algae solution was shaken (80 rpm) continuously for 24 hours. Then the solution was centrifuged at 10,000 rpm and the supernatant was collected for electrolytic and phenolic leakage. The electrolytic leakage was measured by measuring the electrical conductivity of the supernatant. The phenolic leakage was measured by measuring the absorbance of the supernatant at 260 nm using a UV-visible spectrophotometer(40).

**2.5.4 Determination of antioxidant enzyme activities**

The antioxidant enzymes whose catalytic activities were determined in the algae cells were catalase (CAT, enzyme commission number 1.11.1.6), superoxide dismutase (SOD, enzyme commission number 1.15.1.1), ascorbate peroxidase (APX, enzyme commission number 1.11.1.11), and peroxidase (POD, enzyme commission number 1.11.1). For enzyme activity analysis, a fresh pellet of algae cells was prepared as discussed above. A 2 ml of freshly prepared ice-cooled enzyme extraction buffer was added to this algae pellet and vortex the mixture to get a homogenized solution. The enzyme extraction buffer (100 ml) was prepared using 12.5 ml of 0.8 M phosphate buffer, 250 µl of 0.2 N EDTA solution, 0.0176 gm of ascorbic acid, and make up the solution with ultrapure milli-Q water to 100 ml(41). The homogenized solution was centrifuged at 12,000 rpm for 10 min and collect the supernatant (enzyme extract) for enzyme activity analysis. The SOD enzyme (SOD kit, Sigma-Aldrich) and CAT enzymes (catalase kit, Cayman Chemical) activities were determined using SOD and CAT test kits by following the procedure given in the test kits manual.

For POD enzyme activity analysis, a reaction mixture was added to 0.1 ml of enzyme extract. The reaction mixture was made by mixing 3 ml of 100 mM buffer solution, 0.05 ml of guaiacol solution, and 0.03 ml of $H_2O_2$. The reaction starts as soon as the enzyme extract was added to the reaction mixture. The change in absorbance was measured between 30 s to 3 min at 436 nm(42). Moreover, the time required (Δt) to increase in absorbance by 0.1 was used to calculate the POD enzyme activity by the following expression:

$$SOD\ activity\ (U\ l^{-1}) = \frac{500}{\Delta t} \quad (5)$$

For APX enzyme activity analysis, 1 ml of the reaction mixture was prepared. The reaction mixture was prepared by mixing 50 mM potassium phosphate, 0.1 mM $H_2O_2$, 0.5 mM ascorbate, and 0.1 mM EDTA. The reaction started as soon as 0.1 ml of enzyme extract was added to the reaction mixture. The decrease in absorbance was recorded from 10 s to 30 s at 290 nm. The molar extinction coefficient for APX enzyme activity determination used in the present work is given as 2.8 $mM^{-1}\ cm^{-1}$(43).

**2.6 Data analysis**

The results collected during the present work were replicated at least three times (n ≥ 3). The results in the plots are expressed as mean ± standard deviation (μ ± σ). The statistical analysis of the results was performed using a one-way ANOVA followed by a post-hoc test (Bonferroni test) with a statistically significant level of 95% (p-value of 0.05) among the groups. The different lowercase letters were used to show statistically significant differences among the groups in plots.

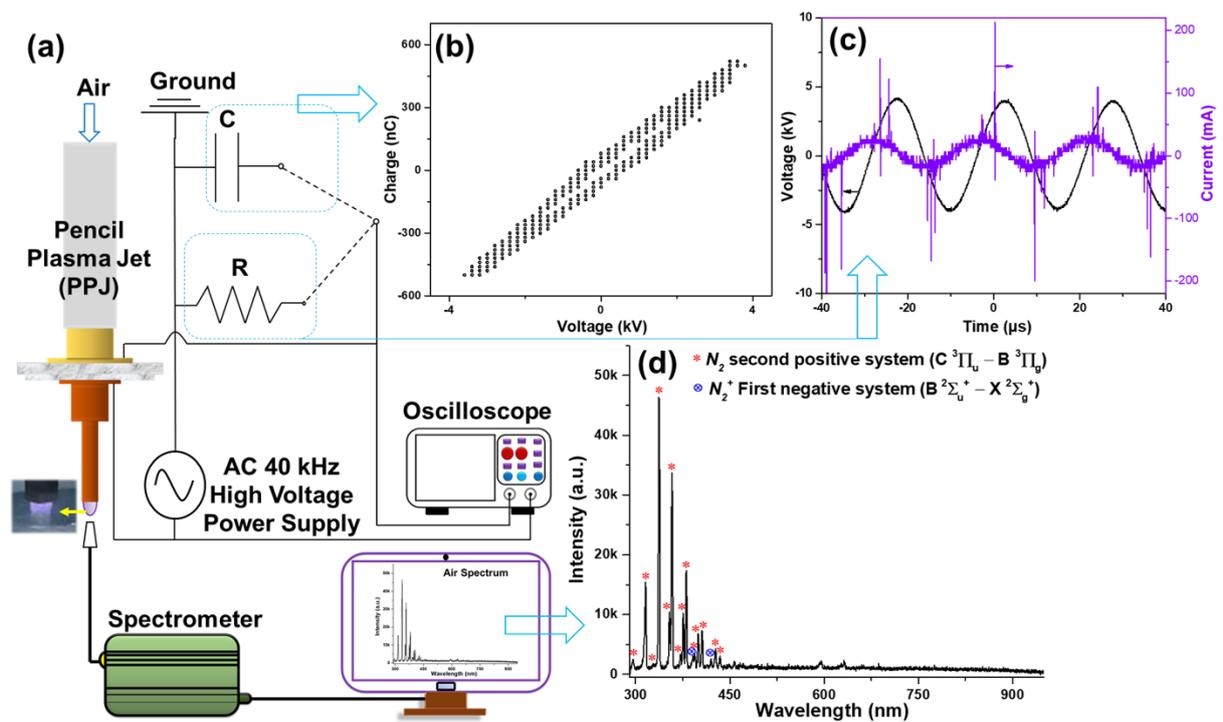

Figure 1. (a) Schematic of electrical and optical emission characterization of pencil plasma jet (PPJ). (b) Voltage-charge Lissajous figure of PPJ, (c) Voltage-current waveform of PPJ, and (d) Optical emission spectrum of air plasma

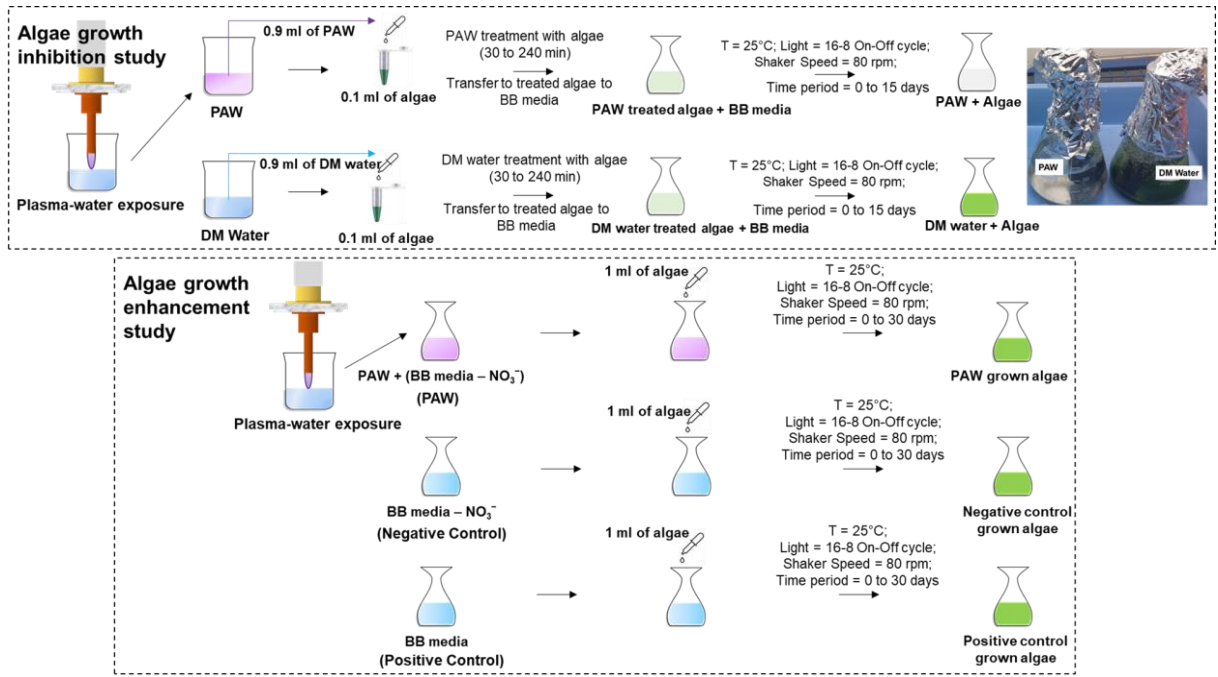

Figure 2. Schematic of algae growth inhibition and algae growth enhancement using plasma activated water.

## 3. Results and discussions

### 3.1 Electrical characterization of DBD-PPJ

The results of the electrical characterization of DBD-PPJ is shown in figure 1 (b, c). In which voltage-charge Lissajous figure and voltage-current waveform is shown. The Lissajous figure (figure 1 (b)) was used to calculate the plasma discharge power or energy consumption during PAW preparation(44). The calculated energy consumption and discharge power were given as 2 mJ and 80 W, respectively. The consumed power results in the generation of various reactive species and radicals in the plasma phase. These reactive species and radicals come in contact with water and form various reactive oxygen-nitrogen species in it.

The voltage-current waveform shown in figure 1 (c) showed the plasma characteristic. The current was a combination of alternating current (AC) and discharge current. The discharge current showed the current flow through new generated species and radicals (45). The observed current peaks in figure 1 (c) in rising and falling half-cycles showed the discharge current. The discharge current represents the discharge of accumulated charge on the dielectric surface on rising and falling current half-cycles. The discharge current peaks were the combination of nanosecond current filaments pulses. These nanosecond pulses combine to give discharge in the microsecond range. Hence, the observed discharge in PPJ is DBD filamentary micro-discharge in nature(44).

## 3.2 Optical emission characterization of DBD-PPJ

The emission spectrum of air plasma is shown in figure 1 (d). The air spectrum showed strong emission band peaks of the $N_2$ second positive system (C $^3\Pi_u \rightarrow$ B $^3\Pi_g$) and weak emission band peaks of the $N_2^+$ first negative system (B $^2\Sigma_u^+ \rightarrow$ X $^2\Sigma_g^+$). The ground state $N_2$ molecule (from air) (X $^1\Sigma_g^+$) goes to $N_2$ upper level (C $^3\Pi_u$) by direct electron impact excitation (E > 11.1 eV) in the plasma phase. Hence, the radiative decay of $N_2$ upper level (C $^3\Pi_u$) to lower $N_2$ level (B $^3\Pi_g$) results in the formation of $N_2$ second positive system. Similarly, the ground state $N_2$ molecule (X $^1\Sigma_g^+$) goes to $N_2^+$ upper level (B $^2\Sigma_u^+$) by electron impact ionization (E > 18.7 eV) in the plasma phase. The radiative decay of this upper state $N_2^+$ (B $^2\Sigma_u^+$) to ground state $N_2^+$ (X $^1\Sigma_g^+$) results in the formation of $N_2^+$ first negative system observed in air emission spectra.

The observed band peaks of $N_2$ second positive systems ($\upsilon''\rightarrow\upsilon'$) were given as 296.1nm (3→1), 315.6 nm (3→1), 327.8 nm (3→3), 337.1 nm (0→0), 353.6 nm (1→2), 357.7 nm (0→1), 371.4 nm (2→4), 375.5 nm (1→3), 380.5 nm (0→2), 394.3 nm (2→5), 399.4 nm (1→4), 405.7 nm (0→3), 427.2 nm (1→5), 434.4 nm (0→4), respectively. Similarly, the observed band peaks of the $N_2^+$ first negative system ($\upsilon''\rightarrow\upsilon'$) were given as 391.0 nm (0→0) and 419.9 nm (2→3)(46, 47).

## 3.3 Physicochemical properties of PAW and RONS concentration

The discharge current characteristics and emission spectroscopy (figure 1 (b, d)) show high energy reactive species produced during air discharge. These discharge gases species when dissolved in the form of various reactive oxygen-nitrogen species (RONS) and change the physicochemical properties of water.

Initially, in the plasma phase, dissociation of $N_2$, $O_2$, and $H_2O$ (from atmospheric moisture) occurs by high-energy electrons, atoms, and molecules to form atomic and molecular species and radicals. These atomic and molecular species consist of N, O, H, and OH, etc. The dissociated species or radicals combine to form other species and radicals like OH, NO, $O_3$, $HO_2$, and $NO_2$, etc. in the plasma phase. The dissociated species, high-energy neutral species, and newly generated species come from the plasma phase to the plasma-liquid interphase. The plasma-liquid interphase region also plays a significant role in plasma chemistry in the liquid phase. Since, the larger distance between plasma and liquid phase results in a thicker plasma-liquid interface. Hence, the short-lived plasma species and radicals generated in the plasma phase disappear before reaching the liquid phase. This results in a low concentration of dissolved species in water which is formed by short-lived species water interaction such as $H_2O_2$. Since the half-life of ˙OH radical is in nanoseconds (48). Therefore, an appropriate distance between the plasma-liquid interface is necessary during PAW production. In plasma-liquid interphase, a wide variety of energetic species/radicals present such as N, $N_2$, $N_2^+$, O, $O_2$, H, OH, $e^-$, NO, $NO_2$, $HO_2$, $O_3$, $H_2O_2$, etc(9, 11, 12, 15, 49-52).

The above-mentioned species when comes in contact with water results in the formation of various kinds of reactive species in water and results in PAW formation. Some of the identified species were given as $NO_2^-$ ions, $NO_3^-$ ions, $ONOO^-$, $H_2O_2$, OH, NO, and aqueous $O_3$ (dissolved $O_3$), etc. Among them, $NO_3^-$ ions have the highest stability and showed the highest concentration compared to other dissolved species in PAW(9, 10, 12, 15). The other dissolved species are comparatively unstable they react with others to form other stable products such as $NO_3^-$ ions. For example, in a high reactive PAW environment (high oxidizing potential and low pH), $NO_2^-$ ions reacted with dissolved $O_3$ and $H_2O_2$ present in PAW and formed $NO_3^-$ ions(9, 10, 12, 15). It suppresses the $NO_2^-$ ions, $H_2O_2$, and dissolved $O_3$ concentration and increased the $NO_3^-$ ions concentration in PAW. Moreover, the dissolved $H_2O_2$ and dissolved $O_3$ also reacted with each other, which resulted in further suppression of their concentration in PAW(9, 10, 12, 15).

The reaction between the above-mentioned species substantially increases the PAW reactivity. The $NO_2^-$ ions reaction with $H_2O_2$ generated peroxynitrite ions which are well-known for their microbial inactivation efficacy(51, 52). Peroxynitrite is unable in acidic PAW, hence converted to more stable $NO_3^-$ ions. In addition, dissociation of $H_2O_2$ in PAW forms hydroxyl (OH) radicals which have excellent antimicrobial properties(52).

As discussed above, the most stable species in PAW is $NO_3^-$ ions which also have substantially higher concentrations compared to other dissolved species in PAW. Hence, PAW is also a rich source of nitrogen which open enormous possibilities for use of PAW as a nitrogen source in numerous applications such as in agriculture as a nitrogen fertilizer and chemical industries for the production of nitrogen acids (nitric and nitrous acids), etc.(22, 23, 25)

Hence, PAW has the potential to be used in microbial inactivation and as well as a nitrogen source for numerous applications. The present work also explores the PAW algal growth inhibition and enhancement ability on freshwater algae *Chlorella Pyrenoidosa and Chlorella Sorokiniana*.

The physicochemical properties and RONS concentration in PAW used for algae growth enhancement and inhibition study are shown in figure 3. The PAW prepared for algae growth enhancement and inhibition study were significantly ($p < 0.05$) different from each other. For the algae growth inhibition study, a higher reactive PAW was prepared compared to the PAW prepared for the growth enhancement study. The reactivity of PAW is defined based on its oxidizing tendency and pH. The high reactive PAW has high oxidation-reduction potential (ORP) and low pH. Figure 3 (a, b) showed high reactive PAW (growth inhibition) had low pH, high ORP, TDS, and EC compared to low reactive PAW (growth enhancement). Similarly, the high reactive PAW had higher $NO_3^-$ ions, $NO_2^-$ ions, dissolved $O_3$, and $H_2O_2$ concentrations compared to low reactive PAW (figure 3 (c, d)).

The pH and ORP of PAW used for algae growth enhancement and inhibition were given as 4.5 and 450 mV (growth enhancement), and 2.3 and 650 mV (growth inhibition), respectively. Similarly, the TDS and EC of PAW used for algae growth enhancement and inhibition were given as 350 ppm and 720 µS cm$^{-1}$ (growth enhancement), and 720 ppm and 7000 µS cm$^{-1}$ (growth inhibition). The measured $NO_3^-$ and $NO_2^-$ ions concentration in PAW prepared for algae growth enhancement and inhibition were given as 180 mg l$^{-1}$ and 1.5 mg l$^{-1}$ (growth enhancement), and 490 mg l$^{-1}$ and 8 mg l$^{-1}$ (growth inhibition). And, the measured $H_2O_2$ and dissolved $O_3$ concentrations in PAW prepared for algae growth enhancement and inhibition were given as 1.3 mg l$^{-1}$ and 0.45 mg l$^{-1}$ (growth enhancement), and 1.5 mg l$^{-1}$ and 15 mg l$^{-1}$ (growth inhibition).

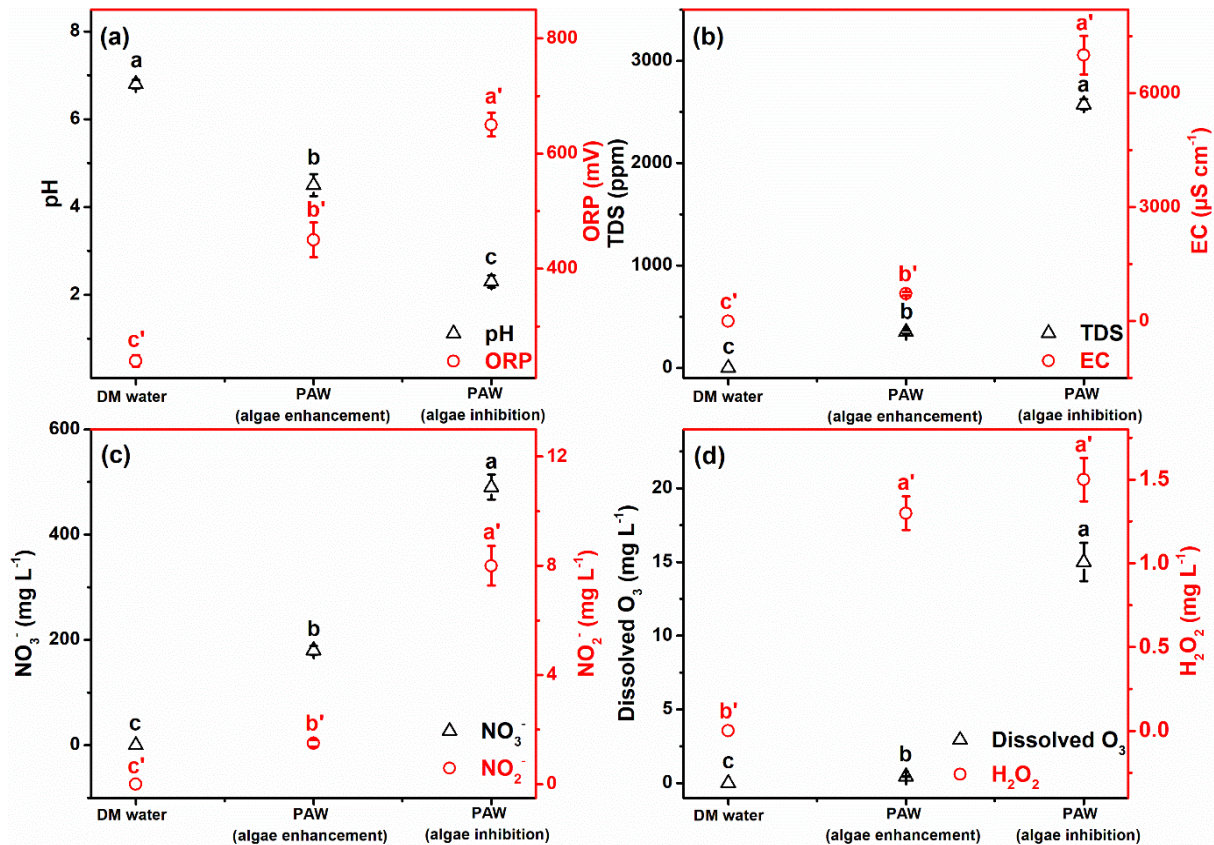

Figure 3. (a, b) The physicochemical properties and (c, d) Reactive oxygen-nitrogen species concentration in plasma activated water when used for algae growth enhancement and inhibition. The statistically significant difference ($p < 0.05$, $n \geq 3$) between the groups is shown by different lowercase letters.

## 3.4 Algae growth inhibition study using PAW

The effect of PAW on the inhibition of freshwater algae *Chlorella Pyrenoidosa* (*C. Pyrenoidosa*) and *Chlorella Sorokiniana* (*C. Sorokiniana*) growth are shown in figure 4. The four different PAW treatment times (30 min, 60 min, 120 min, and 240 min) with algae were studied to study the effect of PAW treatment time on algae growth. The growth of algae, grown using PAW and control were monitored regularly till an incubation period of 15 days as shown in figure 4. As the PAW treatment time increases, we observed a continuous decrease in *C. Pyrenoidosa* and *C. Sorokiniana* growth. For a PAW treatment time of 30 min to 240 min, the decrease in the growth of *C. Pyrenoidosa* and *C. Sorokiniana* were 67.5% and 76.7% respectively at the end of the incubation period (15 days). Moreover, the decrease in the growth of *C. Pyrenoidosa* and *C. Sorokiniana* after 240 min of PAW treatment compared to control were 81.6% and 89.2% at the end of the incubation period. The result obtained at present is also supported by the previously reported work of Mizoi et al.(5). In which they showed RONS present in PAW provide its algicidal efficacy towards *Chlorella vulgaris*. In conclusion, PAW has the potential to inhibit algae growth. Hence, can be used for algicidal applications.

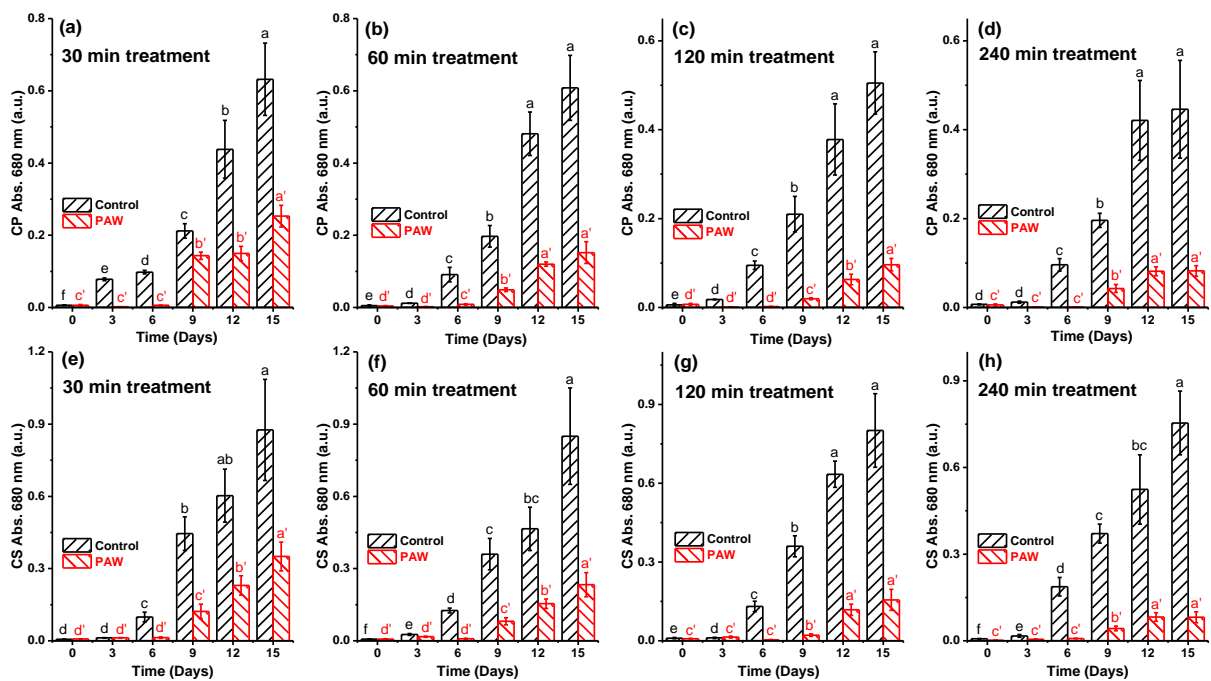

Figure 4. Algae growth inhibition study using plasma activated water (PAW) with varying PAW treatment times. (a-d) *Chlorella Pyrenoidosa,* (e-h) *Chlorella Sorokiniana*. The statistically significant difference ($p < 0.05$, $n \geq 3$) between the groups is shown by different lowercase letters.

### 3.4.1 Morphology analysis of *Chlorella Pyrenoidosa* and *Chlorella Sorokiniana*

The morphology of *C. Pyrenoidosa* and *C. Sorokiniana* after PAW and control treatment is shown in figure 5. The morphology of algae cells was analyzed using high-resolution scanning electron microscopy (HR-SEM). The PAW treated *C. Pyrenoidosa* cells appeared ruptured and shrink. Moreover, the control treated *C. Pyrenoidosa* cells looked healthy and their membrane integrity remains intact (figure 5 (a, b)). Similar results were observed in the morphology of *C. Sorokiniana*, in which PAW treated *C. Sorokiniana* cells appeared damaged and ruptured. However, the control treated *C. Sorokiniana* cells looked healthy and free from any visible damage.

As discussed, the presence of reactive oxygen species such as $H_2O_2$ and dissolved $O_3$, etc. make PAW a powerful oxidizer. Hence, when comes in contact with algae cells, it damaged (oxidize) their membrane which results in leakage of intracellular material enclosed by the membrane. That results in algae cells inactivation which is shown in the results of morphology analysis (figure 5) and grown inhibition plots (figure 4).

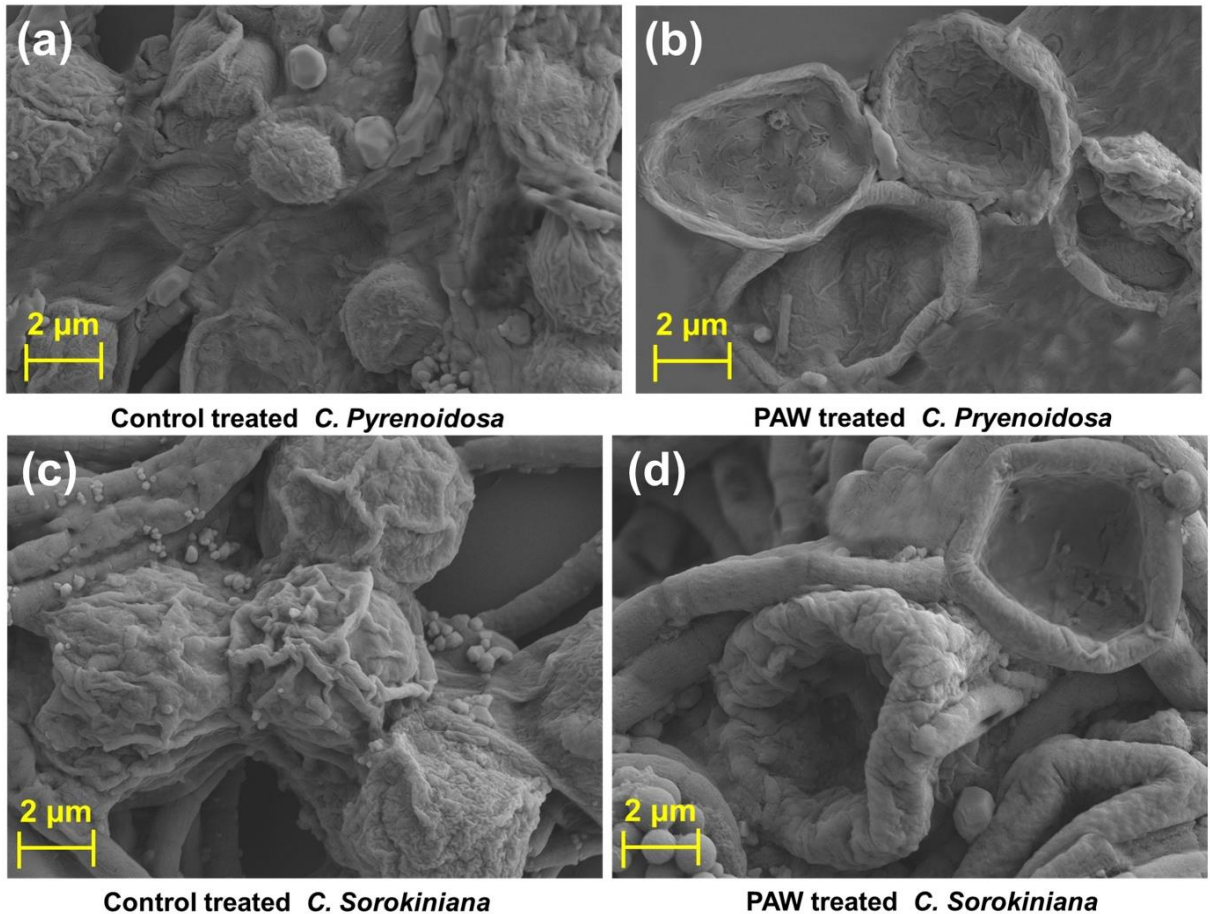

Figure 5. Morphology (SEM images) analysis of *Chlorella Pyrenoidosa* and *Chlorella Sorokiniana* after PAW and control treatment

## 3.5 Algae growth enhancement study using PAW

The role of PAW as a nitrogen source (BB Medium – $NO_3^-$ ions + PAW) in algae growth enhancement (2, 4) compared to a positive control (BB Medium) and negative control (BB medium – $NO_3^-$ ions) is shown in figure 6. Figure 6 (a, b) showed the increase in *C. Pyrenoidosa* and *C. Sorokiniana* growth with time. The PAW grown *C. Pyrenoidosa* showed substantial ($p < 0.05$) higher growth compared to negative and positive control grown *C. Pyrenoidosa*. The increase in *C. Pyrenoidosa* growth with time compared to negative and positive control was given as 200% and 626.3% (day 3), 271.7% and 355.1% (day 6), 623.3% and 217.6% (day 9), 614.3% and 89.6% (day 12), and 579.3% and 17.8% (day 15), respectively.

The *C. Sorokiniana* grown using positive control and PAW as medium showed significantly ($p < 0.05$) higher growth compared to the negative control. Moreover, up to day 6, we observed higher growth in PAW grown *C. Sorokiniana* compared to positive control. The increase in *C. Sorokiniana* growth compared to negative and positive control was given as 105.3% and 35.9% (Day 3), and 249.1% and 102.0% (Day 6). After day 6, PAW and positive control grown *C. Sorokiniana* showed similar growth (not statistically significant difference, $p > 0.05$) with time. Previously reported work of Sukhani et al.(2, 4) suggested that PAW can be used as a nitrogen source in the growth medium for mixed culture algae growth. However, the observed mixed culture algae growth was not significantly higher compared to control used. Hence, based on the above discussion the use of PAW as a nitrogen source has the potential to be used in various agriculture and aquaculture applications.

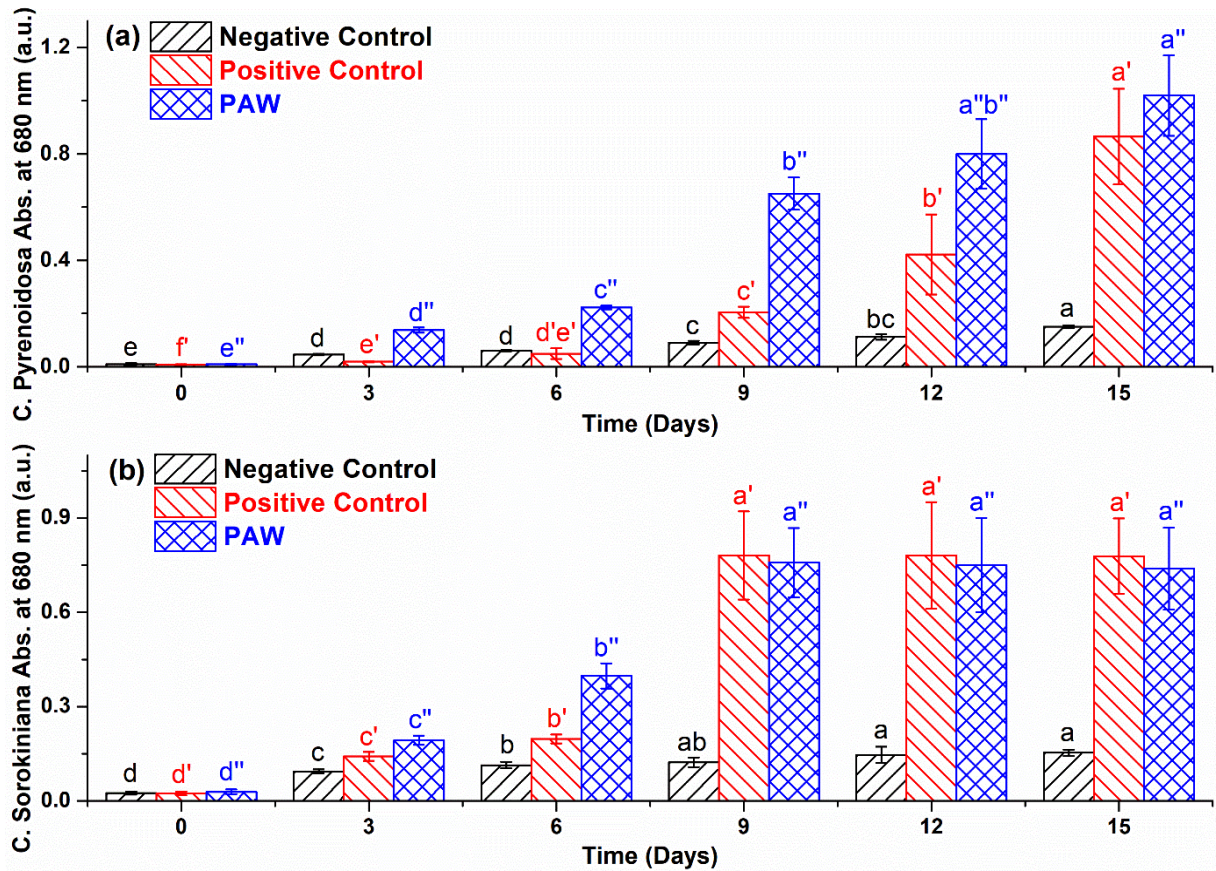

Figure 6. Algae growth enhancement study using PAW and control on (a) *Chlorella Pyrenoidosa* and (b) *Chlorella Sorokiniana*. Negative Control: BB medium – $NO_3^-$ ions, Positive control: BB medium, PAW: BB medium – $NO_3^-$ ions + PAW. The statistically significant difference ($p < 0.05$, $n \geq 3$) between the groups is shown by different lowercase letters.

### 3.5.1 Chlorophyll and Carotenoids concentration

The variation in chlorophyll, carotenoids, soluble sugar, and soluble protein of PAW and control (negative and positive) grown *C. Sorokiniana and C. Pyrenoidosa* is shown in figure 7. The concentration of chlorophyll '$C_a$', chlorophyll '$C_b$', and carotenoids '$C_{x+c}$' in *C. Sorokiniana* were significantly higher when PAW was used as a growth medium compared to negative and positive control (figure 7 (a)). The increase in chlorophyll '$C_a$', chlorophyll '$C_b$', and carotenoids '$C_{x+c}$' were 294.9% and 48.8% (chlorophyll '$C_a$'), 233.3% and 200.0% (chlorophyll '$C_b$'), and 212.6% and 21.9% (carotenoids '$C_{x+c}$'), compared to negative and positive control. The ratio of chlorophyll '$C_a C_b^{-1}$' in *C. Sorokiniana* was substantially higher when positive control was used as growth medium compared to negative control and PAW. It was 138.8% and 101.6% higher than negative control and PAW. Hence, *C. Sorokiniana* grown using positive control as a growth medium have higher photosynthetic capacity compared to negative control and PAW.

Moreover, the PAW grown *C. Pyrenoidosa* showed higher chlorophyll '$C_a$' and chlorophyll '$C_b$' compared to negative and positive control (figure 7 (a)). The percentage increase in chlorophyll '$C_a$' and chlorophyll '$C_b$' in *C. Pyrenoidosa* using PAW as growth medium were given as 170.0% and 14.4% (chlorophyll '$C_a$') and 1098.0% and 35.5% (chlorophyll '$C_b$') compared to negative and positive control. However, chlorophyll '$C_a$' and chlorophyll '$C_b$' concentration of PAW and positive control grown *C. Pyrenoidosa* was not statistically significant ($p > 0.05$). The *C. Pyrenoidosa* grown using positive control showed 116.1% and 13.0% higher carotenoids '$C_{x+c}$' concentrations compared to negative control and PAW. Similar to chlorophyll concentration, the carotenoids '$C_{x+c}$' concentration in *C. Pyrenoidosa* grown using PAW and positive control as growth medium was insignificant ($p > 0.05$). The observed ratio of chlorophyll '$C_a C_b^{-1}$' of negative control grown *C. Pyrenoidosa* was 274.8% and 343.9% higher compared to positive control and PAW. This higher chlorophyll '$C_a C_b^{-1}$' in *C. Pyrenoidosa* grown using negative control as a growth medium signifies its higher photosynthetic potential compared to positive control and PAW.

The higher concentration of chlorophyll '$C_a$' showed more light harvest in Photosystem I and Photosystem II reaction centers resulting in higher oxygenic photosynthesis (53, 54). Hence, the PAW grown *C. Sorokiniana and C. Pyrenoidosa* showed more oxygenic photosynthesis compared to negative and positive control. The higher concentration of chlorophyll '$C_b$' expends the light-absorbing wavelength of algae. The captured energy is transferred to chlorophyll '$C_a$' for oxygenic photosynthesis. Similar to chlorophyll '$C_b$', carotenoids ($C_{x+c}$) also trap the light energy and pass it to chlorophyll '$C_a$' for oxygenic photosynthesis. Hence, chlorophyll '$C_b$' and carotenoids ($C_{x+c}$) act as accessory pigments for oxygenic photosynthesis. In addition, the greenness ratio (('$C_a$' + '$C_b$')/'$C_{x+c}$') of *C. Sorokiniana and C. Pyrenoidosa* grown using PAW as a growth medium was higher compared to negative and positive control.

### 3.5.2 Soluble sugar and protein concentration

The observed variation in soluble protein and soluble sugar of PAW and control grown algae is shown in figure 7 (b). The *C. Sorokiniana and C. Pyrenoidosa* grown using PAW as the growth medium showed higher sugar concentrations compared to negative control and positive control. The higher concentration of soluble sugar in *Pyrenoidosa* and *C. Sorokiniana* grown using PAW showed a better structural and metabolic state of algal cells compared to negative and positive control.

Similar to sugar, the *C. Sorokiniana and C. Pyrenoidosa* grown using PAW as growth medium showed higher soluble protein compared to negative and positive control. However, *C. Sorokiniana* protein grown in PAW was not statistically ($p > 0.05$) significant compared to the positive and negative controls. The higher soluble protein concentration in *C. Sorokiniana and C. Pyrenoidosa* grown using PAW showed better structural, enzymatic, and functionality compared to control. In conclusion, the *C. Sorokiniana and C. Pyrenoidosa* grown using PAW as growth medium have higher soluble protein and sugar which implies better biosynthesis, transportation, and immunity, etc. compared to control.

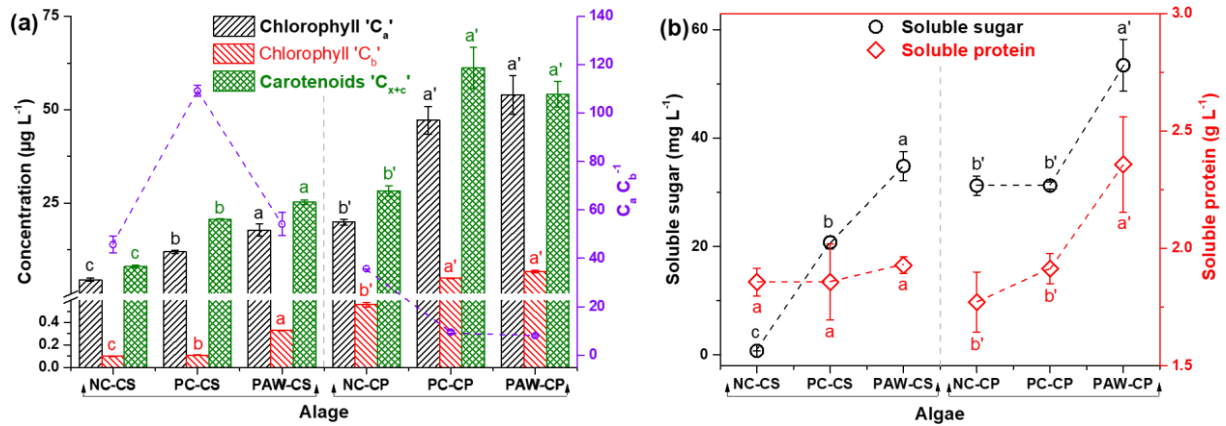

Figure 7. (a) Chlorophyll and carotenoids analysis of PAW and control grown algae, (b) Soluble sugar and soluble protein analysis of PAW and control grown algae. NC: Negative Control (BB medium – $NO_3^-$ ions), PC: Positive Control (BB medium), PAW: BB medium – $NO_3^-$ ions + PAW, CP: *Chlorella Pyrenoidosa*, CS: *Chlorella Sorokiniana*. The statistically significant difference ($p < 0.05$, $n \geq 3$) between the groups is shown by different lowercase letters.

### 3.5.3 Hydrogen peroxide, electrolytic leakage, and phenolic leakage

The hydrogen peroxide ($H_2O_2$) concentration, electrolytic and phenolic leakage in PAW and control grown algae is shown in figure 8 (a). The *C. Sorokiniana* grown using PAW and control as growth mediums do not show significant differences in $H_2O_2$ concentration. However, PAW grown *C. Pyrenoidosa* had a slightly elevated $H_2O_2$ concentration level compared to negative and positive control. Hence, *C. Sorokiniana* and *C. Pyrenoidosa* grown using PAW and control did not show substantial differences in the oxidative stress in the form of $H_2O_2$ concentration. However, $H_2O_2$ also acts as a signaling molecule that helps in algae growth enhancement. Hence, a slightly elevated $H_2O_2$ level in PAW grows *C. Pyrenoidosa* helps in *C. Pyrenoidosa* growth.

The electrolytic and phenolic leakage signifies inorganic ions and phenolic compounds leakage from the cell membrane due to compromise in the membrane integrity. The electrolytic leakage measures external stress on the algae cells. Moreover, the phenolic compounds provide resistance towards biotic and abiotic stress. The observed phenolic leakage in *C. Sorokiniana* and *C. Pyrenoidosa* grown using positive control as the growth medium was significantly lower compared to negative control and PAW. Moreover, there was no observable difference between phenolic leakage of negative control and PAW grown *C. Sorokiniana* and *C. Pyrenoidosa* (figure 8 (a)).

The observed electrolytic leakage in *C. Sorokiniana* and *C. Pyrenoidosa* grown using PAW as the growth medium was significantly ($p < 0.05$) lower than the positive control. Hence, the oxidative stress created in PAW grown *C. Sorokiniana* and *C. Pyrenoidosa* much lower than the positive control. In conclusion, PAW and control grown *C. Sorokiniana* and *C. Pyrenoidosa* do not show external oxidative stress which leads to compromise in the membrane integrity of cell membranes. Therefore, the *C. Sorokiniana* and *C. Pyrenoidosa* cells structural integrity remains intact when using PAW and control as growth medium.

### 3.5.4 Antioxidant enzyme activities

The antioxidant enzyme activity of PAW and control grown *C. Sorokiniana* and *C. Pyrenoidosa* is shown in figure 8 (b). The *C. Sorokiniana* and *C. Pyrenoidosa* grown using PAW and positive control showed a substantial ($p < 0.05$) low concentration of superoxide dismutase (SOD) enzyme compared to the negative control. The SOD enzyme acts as the first line of defense mechanism against oxidative stress created by superoxide ($O_2^-$) ions. It accelerates the conversion of $O_2^-$ ions to $O_2$ and $H_2O_2$ (34). The generation of oxidative stress in algal cells occurs as a response to changes in external surroundings. The low level of SOD enzyme activity in *C. Sorokiniana* and *C. Pyrenoidosa* grown using PAW and positive control showed low oxidative stress in them compared to negative control.

The catalase (CAT) enzyme activity in *C. Sorokiniana* grown using PAW and positive control as growth medium were higher than the negative control. However, PAW grown *C. Sorokiniana* did not show any significant ($p < 0.05$) difference between negative control and positive control. Moreover, PAW and positive control grown *C. Pyrenoidosa* had lower CAT enzyme activity compared to negative control. The production of $H_2O_2$ in cells due to oxidative stress was regulated by the CAT enzyme (34). The CAT enzyme catalyzes the dissociation of $H_2O_2$ to $O_2$ and $H_2O_2$. Hence, higher CAT enzyme activity signifies a higher concentration of $H_2O_2$ in cells. The *C. Sorokiniana* and *C. Pyrenoidosa* grown using PAW either had lower CAT activity or insignificant ($p > 0.05$) compared to control showed lesser oxidative stress created by $H_2O_2$ in algal cells grown using PAW.

The ascorbate peroxidase (APX) enzyme reduces the $H_2O_2$ concentration in cells (34). It is present in the $H_2O_2$ scavenging organelles of cells. Hence, it regulates the intracellular reactive oxygen species levels in cells. The *C. Sorokiniana* grown using PAW showed higher APX enzyme activity compared to negative and positive control. In addition, the PAW grown *C. Pyrenoidosa* showed lower APX enzyme activity compared to negative control. The elevated APX enzyme activity of PAW grown *C. Sorokiniana* showed higher oxidative stress which had been regulated by the APX enzyme. In addition, no observable difference in $H_2O_2$ concentration in PAW grown *C. Sorokiniana* compared to control showed $H_2O_2$ levels in cells were regulated.

Peroxidase (POD) enzyme acts as a defense against cell pathogens. It is a heme (electron donor) protein. It mainly catalyzes the oxidation of organic and inorganic compounds (55). The *C. Sorokiniana* and *C. Pyrenoidosa* grown using PAW and control did not show any POD activity as shown in figure 8 (b). Hence, the POD enzyme activity in *C. Sorokiniana* and *C. Pyrenoidosa* was beyond the detection limit of the present investigation. Or, the no observable POD activity signifies limiting pathogen influence on cells due to which defense response of POD enzyme activity is not measurable.

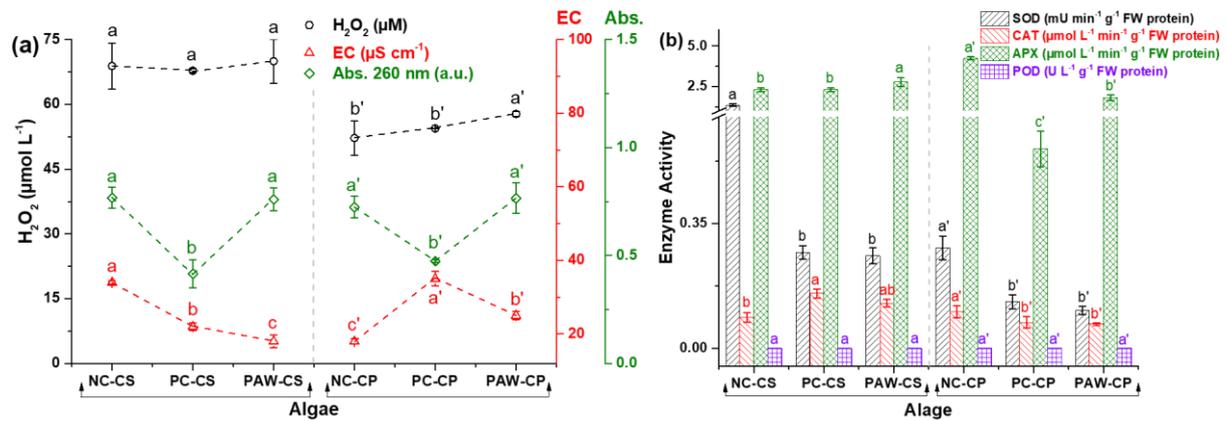

Figure 8. (a) Hydrogen peroxide ($H_2O_2$), electrolytic (EC) and phenolic leakage in PAW and control grown algae, (b) Antioxidant enzyme (SOD, CAT, APX, and POC) activity in PAW and control grown algae. NC: Negative Control (BB medium – $NO_3^-$ ions), PC: Positive Control (BB medium), PAW: BB medium – $NO_3^-$ ions + PAW, CP: *Chlorella Pyrenoidosa*, CS: *Chlorella Sorokiniana*. The statistically significant difference ($p < 0.05$, $n \geq 3$) between the groups is shown by different lowercase letters.

**3.5 Conclusion**

The present study reveals the ability of PAW for algae growth inhibition as well as enhancement. To investigate the algicidal efficacy of PAW a high oxidizing PAW is prepared. The prepared PAW shows a substantial growth inhibition on freshwater algae *C. Sorokiniana* and *C. Pyrenoidosa*. Further, the morphology analysis shows PAW treatment damages and ruptures the *C. Sorokiniana* and *C. Pyrenoidosa* cells membrane which leads to their growth inhibition.

The PAW's role in algae growth enhancement is also investigated. A low oxidizing PAW is used as a nitrogen replacement in Bold's Basal medium. The results are compared with negative control (Bold's Basal medium without nitrogen) and positive control (Bold's Basal medium). The PAW grown *C. Pyrenoidosa* shows a significantly higher growth compared to positive and negative control. Moreover, *C. Sorokiniana* grown using PAW initially showed higher growth compared to negative and positive control and became comparable with positive control later with time. In addition, *C. Sorokiniana* and *C. Pyrenoidosa* grown using PAW as growth medium indicated higher chlorophyll 'a' and 'b', sugar, and protein compared to control. Even with the enhanced growth and elevated nutritional values in PAW grown *C. Sorokiniana* and *C. Pyrenoidosa* cells, they maintain their membrane integrity shown by the results of oxidative stress, and electrolytic and phenolic leakage. Also, the low antioxidant enzymatic activity in PAW grown *C. Sorokiniana* and *C. Pyrenoidosa* signifies the low level of oxidative stress in PAW grown algal cells.

In conclusion, PAW can be prepared with specific physicochemical properties so that it can be used for antialgal activities and as a rich source of nitrogen for its use in agriculture and aquaculture sector (for plants and algal growth enhancement, etc.).

## Acknowledgments

This work was supported by the Department of Atomic Energy (Government of India) graduate fellowship scheme (DGFS). The authors sincerely thank Mr. Vivek and Mr. Sebin for providing constant support and useful suggestions during cells morphology analysis.

## Data availability statement

The data that support the findings of this study are available upon reasonable request from the authors.

## Conflict of interests

The authors declare that there are no conflicts of interests.

## Authors' contributions

Both authors contributed to the study conception and design. Material preparation, data collection, and analysis were performed by Vikas Rathore. The first draft of the manuscript was written by Vikas Rathore, and both authors commented on previous versions of the manuscript. Both authors read and approved the final manuscript.

**ORCID iDs**

Vikas Rathore https://orcid.org/0000-0001-6480-5009
**References**

1. Daneshvar E, Zarrinmehr MJ, Hashtjin AM, Farhadian O, Bhatnagar A. Versatile applications of freshwater and marine water microalgae in dairy wastewater treatment, lipid extraction and tetracycline biosorption. Bioresource technology. 2018;268:523-30.

2. Sukhani S, Punith N, Ekatpure A, Salunke G, Manjari M, Harsha R, et al. Plasma-Activated Water as Nitrogen Source for Algal Growth: A Microcosm Study. IEEE Transactions on Plasma Science. 2021;49(2):551-6.

3. Almarashi JQ, El-Zohary SE, Ellabban MA, Abomohra AE-F. Enhancement of lipid production and energy recovery from the green microalga Chlorella vulgaris by inoculum pretreatment with low-dose cold atmospheric pressure plasma (CAPP). Energy Conversion Management. 2020;204:112314.

4. Sukhani S, Punith N, Lakshminarayana R, Chanakya H. Plasma activated water as a source of nitrogen for algae growth. Advanced Materials Letters. 2019;10(11):797-801.

5. Mizoi K, Rodríguez-González V, Sasaki M, Suzuki S, Honda K, Ishida N, et al. Interactions between pH, reactive species, and cells in plasma-activated water can remove algae. RSC advances. 2022;12(13):7626-34.

6. Heisler J, Glibert PM, Burkholder JM, Anderson DM, Cochlan W, Dennison WC, et al. Eutrophication and harmful algal blooms: a scientific consensus. Harmful algae. 2008;8(1):3-13.

7. Li B, Yu Y, Ye M. Effects of plasma activated species produced by a surface micro-discharge device on growth inhibition of cyanobacteria. Plasma Research Express. 2019;1(1):015017.

8. Corella Puertas E, Dzafic A, Coulombe S. Investigation of the electrode erosion in pin-to-liquid discharges and its influence on reactive oxygen and nitrogen species in plasma-activated water. Plasma Chemistry Plasma Processing. 2020;40(1):145-67.

9. Rathore V, Nema SK. Optimization of process parameters to generate plasma activated water and study of physicochemical properties of plasma activated solutions at optimum condition. Journal of Applied Physics. 2021;129(8):084901.

10. Rathore V, Nema SK. A comparative study of dielectric barrier discharge plasma device and plasma jet to generate plasma activated water and post-discharge trapping of reactive species. Physics of Plasmas. 2022;29(3):033510.

11. Rathore V, Nema SK. The role of different plasma forming gases on chemical species formed in plasma activated water (PAW) and their effect on its properties. Physica Scripta. 2022.

12. Rathore V, Patil C, Sanghariyat A, Nema SK. Design and development of dielectric barrier discharge setup to form plasma-activated water and optimization of process parameters. The European Physical Journal D. 2022;76(5):1-14.

13. Hoeben W, Van Ooij P, Schram D, Huiskamp T, Pemen A, Lukeš PJPC, et al. On the possibilities of straightforward characterization of plasma activated water. Plasma Chemistry Plasma Processing. 2019;39(3):597-626.

14. Raud S, Raud J, Jõgi I, Piller C-T, Plank T, Talviste R, et al. The production of plasma activated water in controlled ambient gases and its impact on cancer cell viability. Plasma Chemistry Plasma Processing. 2021;41(5):1381-95.

15. Rathore V, Patel D, Butani S, Nema SKJPC, Processing P. Investigation of physicochemical properties of plasma activated water and its bactericidal efficacy. Plasma Chemistry Plasma Processing. 2021;41(3):871-902.

16. Rathore V, Patel D, Shah N, Butani S, Pansuriya H, Nema SKJPC, et al. Inactivation of Candida albicans and lemon (Citrus limon) spoilage fungi using plasma activated water. Plasma Chemistry Plasma Processing. 2021;41(5):1397-414.

17. Pan J, Li Y, Liu C, Tian Y, Yu S, Wang K, et al. Investigation of cold atmospheric plasma-activated water for the dental unit waterline system contamination and safety evaluation in vitro. Plasma Chemistry Plasma Processing. 2017;37(4):1091-103.

18. Ten Bosch L, Köhler R, Ortmann R, Wieneke S, Viöl W. Insecticidal effects of plasma treated water. International journal of environmental research public health. 2017;14(12):1460.

19. Guo J, Huang K, Wang X, Lyu C, Yang N, Li Y, et al. Inactivation of yeast on grapes by plasma-activated water and its effects on quality attributes. Journal of food protection. 2017;80(2):225-30.

20. Guo L, Yao Z, Yang L, Zhang H, Qi Y, Gou L, et al. Plasma-activated water: An alternative disinfectant for S protein inactivation to prevent SARS-CoV-2 infection. Chemical Engineering Journal. 2021;421:127742.

21. Xiang Q, Fan L, Li Y, Dong S, Li K, Bai Y. A review on recent advances in plasma-activated water for food safety: Current applications and future trends. Critical Reviews in Food Science Nutrition. 2022;62(8):2250-68.

22. Rathore V, Tiwari BS, Nema SK. Treatment of Pea Seeds with Plasma Activated Water to Enhance Germination, Plant Growth, and Plant Composition. Plasma Chemistry Plasma Processing. 2022;42(1):109-29.

23. Sajib SA, Billah M, Mahmud S, Miah M, Hossain F, Omar FB, et al. Plasma activated water: The next generation eco-friendly stimulant for enhancing plant seed germination, vigor and increased enzyme activity, a study on black gram (Vigna mungo L.). Plasma Chemistry Plasma Processing. 2020;40(1):119-43.

24. Subramanian PG, Rao H, Shivapuji AM, Girard-Lauriault P-L, Rao L. Plasma-activated water from DBD as a source of nitrogen for agriculture: Specific energy and stability studies. Journal of Applied Physics. 2021;129(9):093303.

25. Shainsky N, Dobrynin D, Ercan U, Joshi SG, Ji H, Brooks A, et al. Plasma acid: water treated by dielectric barrier discharge. Plasma Process Polym. 2012;9:1-6.

26. Han S-F, Jin W, Yang Q, Abomohra AE-F, Zhou X, Tu R, et al. Application of pulse electric field pretreatment for enhancing lipid extraction from Chlorella pyrenoidosa grown in wastewater. Renewable Energy. 2019;133:233-9.

27. Eladel H, Abomohra AE-F, Battah M, Mohmmed S, Radwan A, Abdelrahim H. Evaluation of Chlorella sorokiniana isolated from local municipal wastewater for dual application in nutrient removal and biodiesel production. Bioprocess biosystems engineering. 2019;42(3):425-33.

28. Napolitano G, Fasciolo G, Salbitani G, Venditti P. Chlorella sorokiniana dietary supplementation increases antioxidant capacities and reduces ros release in mitochondria of hyperthyroid rat liver. Antioxidants. 2020;9(9):883.

29. Cazzaniga S, Dall'Osto L, Szaub J, Scibilia L, Ballottari M, Purton S, et al. Domestication of the green alga Chlorella sorokiniana: reduction of antenna size improves light-use efficiency in a photobioreactor. Biotechnology for biofuels. 2014;7(1):1-13.

30. Nakano S, Takekoshi H, Nakano M. Chlorella (Chlorella pyrenoidosa) supplementation decreases dioxin and increases immunoglobulin a concentrations in breast milk. Journal of medicinal food. 2007;10(1):134-42.

31. Kholssi R, Marks EA, Miñón J, Montero O, Debdoubi A, Rad C. Biofertilizing effect of Chlorella sorokiniana suspensions on wheat growth. Journal of Plant Growth Regulation. 2019;38(2):644-9.

32. Cavalcanti FR, Oliveira JTA, Martins-Miranda AS, Viégas RA, Silveira JAG. Superoxide dismutase, catalase and peroxidase activities do not confer protection against oxidative damage in salt-stressed cowpea leaves. New Phytologist. 2004;163(3):563-71.

33. Beyer Jr WF, Fridovich I. Assaying for superoxide dismutase activity: some large consequences of minor changes in conditions. Analytical biochemistry. 1987;161(2):559-66.

34. Khan A, Numan M, Khan AL, Lee I-J, Imran M, Asaf S, et al. Melatonin: Awakening the defense mechanisms during plant oxidative stress. Plants. 2020;9(4):407.

35. Murtey MD, Ramasamy P. Sample preparations for scanning electron microscopy–life sciences. Modern electron microscopy in physical life sciences. 2016:161-85.

36. Lichtenthaler HK, Buschmann C. Chlorophylls and carotenoids: Measurement and characterization by UV-VIS spectroscopy. Current protocols in food analytical chemistry. 2001;1(1):F4. 3.1-F4. 3.8.

37. Pons A, Roca P, Aguiló C, Garcia F, Alemany M, Palou AJJob, et al. A method for the simultaneous determinations of total carbohydrate and glycerol in biological samples with the anthrone reagent. Journal of biochemical biophysical methods. 1981;4(3-4):227-31.

38. Lowry O. Rosebrough NJ, Farr Al, and Randall RJ. Protein measurement with the Folin phenol reagent J Biol Chem. 1951;193:265-75.

39. Alexieva V, Sergiev I, Mapelli S, Karanov E. The effect of drought and ultraviolet radiation on growth and stress markers in pea and wheat. Plant, Cell & Environment. 2001;24(12):1337-44.

40. Jain R, Srivastava S, Chandra A. Electrolyte and phenolic leakage contents and their relation with photosynthetic pigments, catalase, peroxidase and superoxide dismutase activities at low temperature stress in sugarcane. Indian J Plant Physiol. 2012;17(3&4):246-53.

41. Almeselmani M, Deshmukh P, Sairam RK, Kushwaha S, Singh T. Protective role of antioxidant enzymes under high temperature stress. Plant science. 2006;171(3):382-8.

42. Saikia D, Kataki MD. COMPARATIVE ENZYME ASSAY OF CARISSA CARANDAS FRUIT AT VARIOUS STAGES OF GROWTH WITH STORED RIPE STAGE. 2014.

43. Nakano Y, Asada K. Hydrogen peroxide is scavenged by ascorbate-specific peroxidase in spinach chloroplasts. Plant cell physiology. 1981;22(5):867-80.

44. Fang Z, Xie X, Li J, Yang H, Qiu Y, Kuffel E. Comparison of surface modification of polypropylene film by filamentary DBD at atmospheric pressure and homogeneous DBD at medium pressure in air. Journal of Physics D: Applied Physics. 2009;42(8):085204.

45. Sato N. Discharge current induced by the motion of charged particles. Journal of Physics D: Applied Physics. 1980;13(1):L3.

46. Qayyum A, Zeb S, Ali S, Waheed A, Zakaullah MJPc, processing p. Optical emission spectroscopy of abnormal glow region in nitrogen plasma. Plasma chemistry plasma processing. 2005;25(5):551-64.

47. Shemansky D, Broadfoot A. Excitation of N2 and N+ 2 systems by electrons—I. Absolute transition probabilities. Journal of Quantitative Spectroscopy Radiative Transfer. 1971;11(10):1385-400.

48. Zhu Y, Serra A, Guo T, Park JE, Zhong Q, Sze SK. Application of nanosecond laser photolysis protein footprinting to study EGFR activation by EGF in cells. Journal of Proteome Research. 2017;16(6):2282-93.

49. Daito S, Tochikubo F, Watanabe T. Improvement of NOx removal efficiency assisted by aqueous-phase reaction in corona discharge. Japanese Journal of Applied Physics. 2000;39(8R):4914.

50. Liu Y, Liu D, Luo S, Sun B, Zhang M, Yang A, et al. 1D fluid model of the interaction between helium APPJ and deionized water. Journal of Physics D: Applied Physics. 2022;55(25):255204.

51. Van Gils C, Hofmann S, Boekema B, Brandenburg R, Bruggeman P. Mechanisms of bacterial inactivation in the liquid phase induced by a remote RF cold atmospheric pressure plasma jet. Journal of Physics D: Applied Physics. 2013;46(17):175203.

52. Roy NC, Pattyn C, Remy A, Maira N, Reniers F. NOx synthesis by atmospheric-pressure N2/O2 filamentary DBD plasma over water: Physicochemical mechanisms of plasma–liquid interactions. Plasma Processes Polymers. 2021;18(3):2000087.

53. Watanabe T, Kobayashi M, Hongu A, Nakazato M, Hiyama T, Murata N. Evidence that a chlorophyll a'dimer constitutes the photochemical reaction centre 1 (P700) in photosynthetic apparatus. FEBS letters. 1985;191(2):252-6.

54. Bjijrn L, Papageorgiou G, Blankenship R, Govind-jee A. viewpoint: Why chlorophyll a. Photosynth Res. 2009;99:85-98.

55. Goud PB, Kachole MSJPs, behavior. Antioxidant enzyme changes in neem, pigeonpea and mulberry leaves in two stages of maturity. Plant signaling behavior. 2012;7(10):1258-62.